\documentclass[pra,twocolumn,superscriptaddress]{revtex4}

\newcommand{\be}{\begin{equation}}
\newcommand{\ee}{\end{equation}}

\newcommand{\bea}{\begin{eqnarray}}
\newcommand{\eea}{\end{eqnarray}}

\usepackage{amsfonts}
\usepackage{amsmath}
\usepackage{bbm}
\usepackage{graphicx}
\usepackage{bbold}
\usepackage{amssymb}
\usepackage{color}
\usepackage{subfigure}
\usepackage[allfiguresdraft]{draftfigure}
\usepackage{hyperref}
\usepackage{lipsum}
\usepackage{multirow}
\usepackage{mathtools}
\usepackage{blindtext}
\usepackage{makecell}

\usepackage{array}
\newcolumntype{?}{!{\vrule width 1.25pt}}

\begin{document}
\title{The Haldane Model with Chiral Edge States \protect \\ using a Synthetic Dimension}

\author{Joel Priestley}
\thanks{\href{mailto:jcp2@hw.ac.uk}{jcp2@hw.ac.uk}}
%The lipsum environment doesn't allow for the email to be put in the bottom left corner, taking the \lipsum[]'s out fixes that
\affiliation{SUPA, Institute of Photonics and Quantum Sciences, Heriot-Watt University, Edinburgh EH14 4AS, United Kingdom}
\author{Gerard Valent\'i-Rojas}
\affiliation{SUPA, Institute of Photonics and Quantum Sciences, Heriot-Watt University, Edinburgh EH14 4AS, United Kingdom}
\author{Patrik \"Ohberg}
\affiliation{SUPA, Institute of Photonics and Quantum Sciences, Heriot-Watt University, Edinburgh EH14 4AS, United Kingdom}

\begin{abstract}
\noindent
We explicitly show that the differences, with respect to the appearance of topological phases, between the traditional Haldane model, which utilises a honeycomb lattice structure, to that of the Haldane model imbued onto a brick-wall lattice geometry, are inconsequential. A proposal is then put forward to realise the Haldane model by exploiting the internal degrees of freedom of atoms as a synthetic dimension. This leads to a convenient platform for the investigation of chiral edge states due to the hard boundaries provided by the hyperfine manifold. We make some cursory comments on the effects of interactions in the system.
\end{abstract}

\maketitle
\section{Introduction}

Haldane's work from 1988 using a `toy' model demonstrated how a non-zero band Chern number, and a therefore non-zero quantised Hall conductance, could be achieved in ``$2D$ graphite" with a net zero magnetic field \cite{Haldane:88}. In the article the author expresses doubts about whether the model is physically realisable, yet some twenty years later it was experimentally achieved in an ultracold Fermi gas \cite{Jotzu:14}. Two of the main signatures of topological phases of matter are edge states and a quantised Hall conductance. The first of these hallmarks has been previously observed in synthetic dimensional systems in a quantum Hall regime \cite{Mancini:15, Stuhl:15}. 

In recent years there has been a growing interest in synthetic dimensions where, for example, the states of the hyperfine manifold of an atom are coherently coupled together, allowing for the construction of a $(D+1)-$dimensional system from one which has $D$ spatial dimensions \cite{Boada:12}. This new method has come with its own plaudits and schemes have been proposed which would realise topological states of matter such as the $4D$ quantum Hall effect \cite{Price:15} and even the $6D$ quantum Hall effect \cite{Petrides:18, Lee:18} have been put forward, as well as more exotic physics such as the Creutz-Hubbard model \cite{Junemann:17}. Using synthetic gauge fields, high effective magnetic field strengths, otherwise experimentally unattainable, have been realised in ultracold atoms to simulate the Harper-Hofstadter model, having an important bearing in the realisation of topological phases \cite{Aidelsburger:13, Miyake:13}. In this work, we propose an experimental scheme using synthetic dimensions to attain the Haldane model, on a bipartite brick-wall lattice with complex next-nearest neighbour tunnelling. 

First, we analyse the differences between the Haldane model on its traditional honeycomb lattice and on a brick-wall lattice to ensure the system retains the desired properties. Next, we propose using a synthetic dimension to create the Haldane model - achieved by the use of Raman-assisted tunnelling, which can imprint complex phases on the wave-function of the atom. In implementing this, one would capture the spirit of Haldane's proposal to generate a non-zero Hall conductance without the appearance of Landau levels. One advantage of using such a synthetic dimension are the hard boundaries. The hyperfine manifold provides an ideal platform for the investigation of chiral edge states \cite{Celi:14}, as opposed to in degenerate atomic gases where the edge of the sample is usually imposed gradually, by harmonic confinement and so not reflecting the abrupt termination such as that found on a sample of graphene. The edge state physics is therefore non-trivial and needs to be considered in its own right \cite{Buchhold:12, Stanescu:09}. Synthetic dimensions on the other hand are manipulated optically, leading to high controllability and flexibility.

According to the bulk-boundary correspondence\cite{Hatsugai:09, Delplace:11}, for a non-zero Chern number one expects to observe edge states and previous studies have predicted their existence for the specific case of the Haldane model \cite{Hao:08}. To identify the topological nature of the Haldane model \cite{Jotzu:14}, a constant force is applied resulting in an orthogonal drift, analogous to a Hall current, was observed. The topology of the band can be explored by inducing Bloch oscillations and mapping out in quasi-momentum space the locations of the Dirac points \cite{Jotzu:14}. Chiral edge states in experimental analogues of graphene have been observed in photonic lattice systems \cite{Rechtsman:13}. Edge states in synthetic dimensional system have been previously reported and the way in which they were detected, using spin-selective imaging, is a valuable experimental asset \cite{Mancini:15, Stuhl:15}. 

This article is structured as follows; in Section II we analyse the effects of deforming the lattice from a honeycomb to a brick-wall lattice. In Section III we put forward our experimental proposal for the Haldane model. Section IV focuses on the much sought-after edge states. Section V discusses the effects of interactions. We conclude with a summary of our proposal and outlook for Chern insulators in synthetic dimensional systems in Section VI.

\section{The Brick-wall Haldane Model}
\begin{figure}
\includegraphics[draft=false,width=\columnwidth]{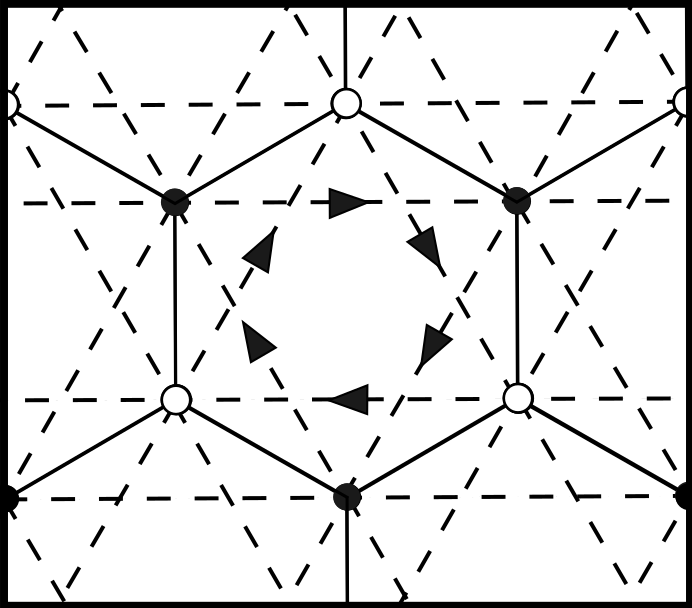}
\includegraphics[draft=false,width=\columnwidth]{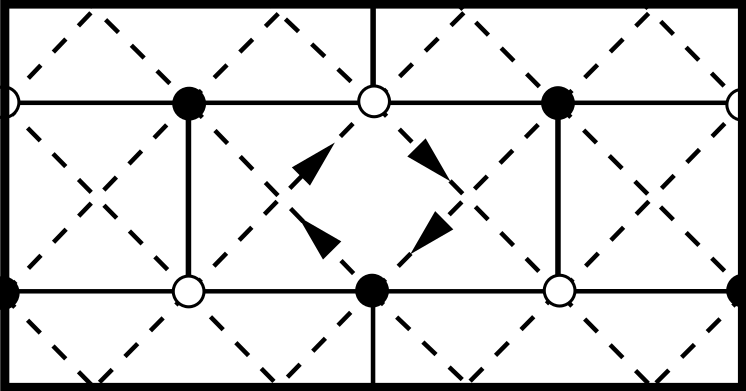}
\caption{(a) [top panel] A honeycomb lattice with next nearest neighbour couplings indicated with dashed lines and with arrows showing the direction of positive phase gain. $A$ and $B$ sites are represented by white and black dots, respectively (b) [bottom panel] Brick-wall lattice, same conventions as top panel.}
\label{lattices}
\end{figure}

The Haldane model \cite{Haldane:88} fundamentally relies on the breaking of two symmetries. The first of which is inversion symmetry, accomplished by making the lattice bipartite. Mathematically, this is represented by an energy offset $\pm M$ between neighbouring sites, labelled respectively as $A$ and $B$. The second symmetry is time-reversal symmetry. This is broken by the introduction of complex next-nearest tunnelling which involves the accumulation of a phase. This can be viewed as a Peierls phase or flux through a plaquette. The competition between these two broken symmetries determines which phase the system is in. The eigenvector of a bipartite Hamiltonian, written in Bloch form, is a two-vector whose elements are the wavefunctions on the $A$ and $B$ sites. The Haldane model Bloch Hamiltonian is given by \eqref{Haldane}
\begin{align}\label{Haldane}
\mathbf{\hat{H}}=\sum_{\mathbf{k}}\mathbf{\hat{\Psi}}^\dagger(\mathbf{k})\mathcal{\hat{H}}(\mathbf{k})\mathbf{\hat{\Psi}}(\mathbf{k})
\end{align}
with
\begin{widetext}
\begin{align}
\mathcal{\hat{H}}(\mathbf{k}) =\left(2t_2\:\text{cos}\:\phi\left[\sum_i\text{cos}(\mathbf{k}\cdot\mathbf{b}_i)\right]\mathbf{I}
+t_1\left(\sum_i[\text{cos}(\mathbf{k}\cdot\mathbf{a}_i)\mathbf{\sigma}_x+\text{sin}(\mathbf{k}\cdot\mathbf{a}_i)\mathbf{\sigma}_y]\right) 
+\left\{M-2t_2\:\text{sin}\:\phi\left[\sum_i \text{sin}(\mathbf{k}\cdot\mathbf{b}_i)\right]\right\}\mathbf{\sigma}_z\right)\nonumber
\end{align}
\end{widetext}
where $\mathbf{\hat{\Psi}}(\mathbf{k})=(\hat{\psi}_A,\hat{\psi}_B)^\top$ is the Bloch state spinor. This form of the Hamiltonian is general, no choice of geometry has yet been made and this is enforced by the input of specific
vectors $\mathbf{a}_i, \mathbf{b}_i$ as the arguments of the trigonometric functions.
Notation-wise, $t_{1}$ is the real nearest-neighbour tunnelling probability, $t_{2}$ is the real part of the next-nearest neighbour tunnelling probability, $\phi$ is the phase gain associated to next-nearest neighbour hopping, mathematically originating from the imaginary part of the next-nearest neighbour tunnelling. Moreover, $\mathbf{a}_i$ and $\mathbf{b}_i$ are the spatial vectors to nearest- and next-nearest neighbours, respectively, and $\sigma_{x,y,z}$ are the Pauli matrices. The spatial vectors from a $B$ site to a nearest neighbour $A$ site for the honeycomb lattice are ${(\pm \frac{a}{2}, \; \frac{\sqrt{3}a}{2})^{\top},( 0, \; -a)^{\top}}$ and the next-nearest neighbour vectors are consequently given by $\mathbf{b}_1=\mathbf{a}_2-\mathbf{a}_3$, $\mathbf{b}_2=\mathbf{a}_3-\mathbf{a}_1$ and $\mathbf{b}_3=\mathbf{a}_1-\mathbf{a}_2$. The brick-wall lattice vectors are given by ${(\pm a, \; 0)^{\top},( 0,\; -a)^{\top}}$ and the next-nearest neighbour displacements as above. In both cases $a$ is the lattice spacing, that is to say the shortest distance between any two sites. To break time-reversal symmetry, phase gain is only positive from next-nearest neighbour hopping in a clockwise circulation as shown in Figure \eqref{lattices}.

\begin{figure}
\includegraphics[draft=false,width=\columnwidth]{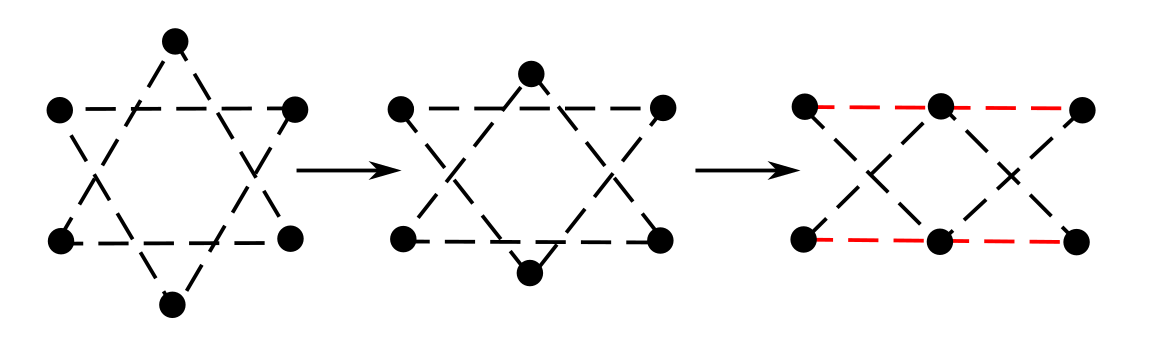}
\caption{Deformation of honeycomb lattice to brick-wall lattice geometry - for demonstrative purposes only next-nearest neighbour links are shown. Red links are the next-nearest neighbour couplings that we consider dropping since they align with nearest neighbour couplings which enclose no flux.}
\label{deform}
\end{figure}

\begin{figure}
\includegraphics[draft=false,width=\columnwidth]{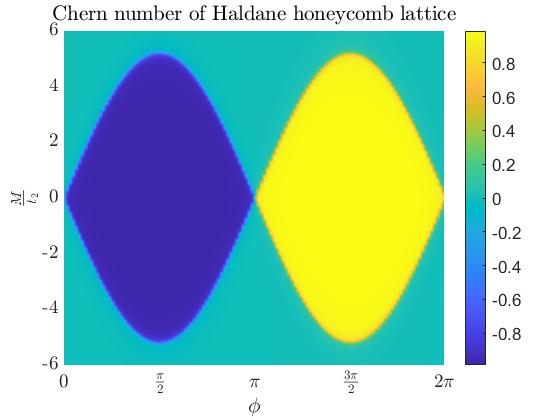}
\includegraphics[draft=false,width=\columnwidth]{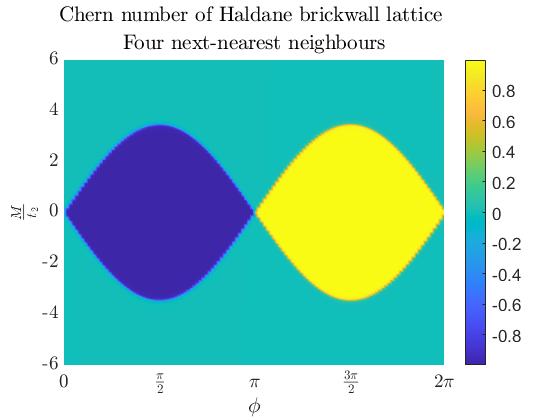}
\caption{(a) [top panel] The phase diagram for the honeycomb lattice. The six next-nearest neighbour brick-wall gives the same plot (b) [bottom panel] The four next-nearest neighbour brick-wall lattice retains the same features but with a reduced area in parameter space for a topologically non-trivial band structure.}
\label{chern}
\end{figure}

The Hamiltonian \eqref{Haldane} can be put in the form
\begin{align}\label{Bloch}
\mathcal{\hat{H}}(\mathbf{k})=\varepsilon(\mathbf{k})\mathbf{I}+\mathbf{d}(\mathbf{k})\cdot\mathbf{\sigma}, 
\end{align}

with $\mathbf{\sigma}$ being the Pauli vector, and the spectrum given by
\begin{align}\label{Spectrum}
E_{\pm}(\mathbf{k})=\varepsilon(\mathbf{k})\pm\sqrt{\mathbf{d}(\mathbf{k})\cdot\mathbf{d}(\mathbf{k})},
\end{align}
where $\varepsilon(\mathbf{k})$ is the energy that both bands share and $\mathbf{d}(\mathbf{k})$ are the coefficients of the Pauli matrices. By first defining the unit vector of these coefficients
\begin{align}\label{Unit}
\mathbf{n}(\mathbf{k}) =\frac{\mathbf{d}(\mathbf{k})}{|\mathbf{d}(\mathbf{k})|},
\end{align}
the topological invariant of an insulator with broken time-reversal symmetry, the first Chern number, can be calculated. The Chern number is related to the solid angle which is subtended by the unit vector $\mathbf{n}(\mathbf{k})$ in $\mathbf{k}$-space as $\mathbf{k}$ runs over the Brillouin zone. Therefore, for the lower band this can be found by \cite{Hasan:10}
\begin{align}\label{Berry}
\nu_1 = -\frac{1}{4\pi} \iint_{\text{BZ}} \text{d}^2 k \; \mathbf{n} \cdot \left[ (\partial_{k_x}\mathbf{n}) \times (\partial_{k_y}\mathbf{n})\right],
\end{align}
where the integrand is known as the Berry curvature and the integral is over the first Brillouin zone. 
Then, the Hall conductance of the system is given by
\begin{align}\label{QHE}
    \sigma_{xy} = \nu_1 \frac{e^2}{h}
\end{align}
despite the net magnetic field $\mathbf{B} = \mathbf{0}$. The Haldane model is therefore a topological insulator.
The brick-wall lattice is topologically equivalent to the honeycomb lattice. One could imagine smoothly deforming the honeycomb lattice by flattening out the top and bottom points of the hexagon. The Dirac points, if the system is in a regime such that these points exist, move away from the boundary of the Brillouin zone at $\mathbf{K} = \frac{2\pi}{3a}(1,\;\frac{1}{\sqrt{3}})^{\top}$ and $\mathbf{K'} = \frac{2\pi}{3a}(1,\;-\frac{1}{\sqrt{3}})^{\top}$ to a point deeper within the, now square, Brillouin zone at $(\pm\frac{2\pi}{3a},\; 0)^{\top}$. The honeycomb lattice with three nearest neighbours has six next-nearest neighbours. It is not unreasonable to consider that the next-nearest neighbour hoppings which go along the long-side of the real space lattice in the brick-wall case can be dropped, see Figure \eqref{deform}, as the sites which this tunnelling connects move away from each other. There are therefore two options in the brick-wall case, six or four next-nearest neighbours. 
To ensure that no relevant physics is lost by changing the lattice from a honeycomb to a brick-wall lattice, the Chern number as a function of phase and $M/t_2$ is calculated numerically, see Figure \eqref{chern}.
The 6 next-nearest neighbour Brick-wall and honeycomb lattice situations are identical. In the 4 next-nearest neighbour brick-wall system, the boundaries of the phases, \textit{i.e.} the lines in parameter space where the phase transition occurs, are moved, relative to the 6 next-nearest neighbour case. A phase transition occurs when a band gap closes and at this Dirac point, non-trivial topology in the form of Berry curvature is introduced to the band structure. Since these Dirac points are a point of degeneracy in the spectrum they can be easily calculated using Equation \eqref{Spectrum}, by finding the points in reciprocal space for which $\mathbf{d}(\mathbf{k})=0$. This leads to boundaries defined by $\frac{M}{t_2}=\pm3\sqrt{3}\:\text{sin}\:\phi$ for the honeycomb and 6 next-nearest neighbour brick-wall lattice models, and $\frac{M}{t_2}=\pm2\sqrt{3}\:\text{sin}\:\phi$ for the 4 next-nearest neighbour brick-wall case. It is instructive to verify that the physics remains invariant, even the lattice is deformed topologically, since there are further symmetries that are broken. The honeycomb lattice enjoys a six-fold rotational symmetry which is reduced to a three-fold with the introduction of an energy off-set, distinguishing every other site. The brick-wall lattice however has a two-fold rotational symmetry which is reduced to the trivial group; that is to say it only contains the identity, when $M \neq 0$. As can be seen from the phase diagram in Figure \eqref{chern} (b), even for $M = 0$ the system is in a topological regime, $\phi\neq 0,\pi$, with $\nu_1 = \pm 1$ depending on the flux.
\subsection*{Anisotropy}
\begin{figure}
\includegraphics[draft=false,width=\columnwidth]{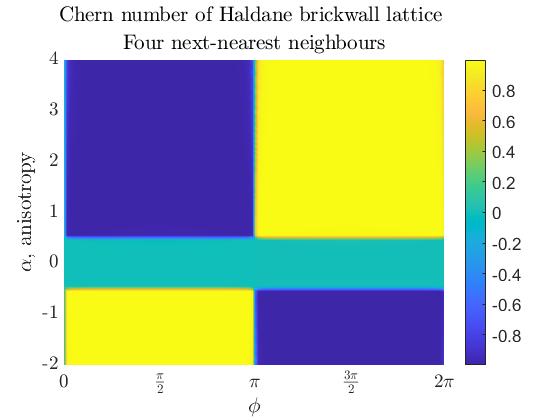}
\caption{Phase diagram for anisotropy in next-nearest neighbour tunnelling versus phase gain, with fixed on-site energy difference $M = 0$.}
\label{anisotropy}
\end{figure}

Next, we investigate the effect of an anisotropy in the coupling constants of the model. We focus on the brick-wall 4 next-nearest neighbours case. This is a suitable candidate for our proposal since it requires fewer lasers as they correspond to next-nearest neighbour hoppings. One can introduce a directional anisotropy by setting $J_y = t_1$ and $J_x = \alpha t_1$, and then by varying $\alpha$ investigate the effect of the topological phases due to the unequal coupling constants. For a vanishing on-site energy, $M = 0$, one can plot a phase diagram of anisotropy versus flux, see Figure \eqref{anisotropy}.
Similar results are obtained for an anisotropy in the $y$-direction. If $M\neq 0$ then there remains a range of $\phi$ for which the Chern number vanishes, as expected.

\newpage
\begin{widetext}
\begin{center}
\begin{table}
\begin{tabular}{??c|c|c??}
\Xhline{1.5pt}
Difference & Honeycomb & Brick-wall \\
\Xhline{1.25pt}
Nearest neighbour & $\begin{pmatrix}\pm \frac{1}{2} \\ \frac{\sqrt{3}}{2} \end{pmatrix}a$, $\begin{pmatrix}0 \\ -1 \end{pmatrix}a$ & $\begin{pmatrix} \pm 1 \\ 0 \end{pmatrix}a$, $\begin{pmatrix}0 \\ -1 \end{pmatrix}a$ \\ 
\hline
Next-nearest neighbour & $\pm(\mathbf{a}_2-\mathbf{a}_3),\pm(\mathbf{a}_3-\mathbf{a}_1),\pm(\mathbf{a}_1-\mathbf{a}_2)$ & $\pm(\mathbf{a}_1+\mathbf{a}_2),\pm(\mathbf{a}_2-\mathbf{a}_1), \pm2\mathbf{a}_1$ \\
\hline
Flux & $\phi = \frac{2\pi(2\Phi_a+\Phi_b)}{\Phi_0}$ & $\phi = \frac{2\pi(\Phi_a+\Phi_b)}{\Phi_0}$ \\
\hline
Brillouin zone & Hexagon & $\frac{\pi}{4}$ rotated square \\
\hline
Dirac points & $M = \pm3\sqrt{3} \:t_2\:\text{sin}\:\phi$ & $M = \pm3\sqrt{3} \:t_2\:\text{sin}\:\phi$ \\
\Xhline{1.5pt}
\end{tabular}
\caption{Summary of differences between the Honeycomb and Brick-wall lattice Haldane models}\label{diff}
\end{table}
\end{center}
\end{widetext}

This allows for differing coupling constants along different axes provided they are not too dissimilar, as this is more likely experimentally attainable. As can be seen from Figure \eqref{anisotropy}, if $|J_y/J_x| > 2$ then the non-zero Chern numbers remain but are sign flipped for negative $\alpha$. Anisotropy has previously been shown to affect the appearance of edge states in graphene. At the same critical point that is found in our results above, $J_y = 2 J_x$, edge states on the zig-zag edge were destroyed due to the merging of Dirac points, annihilating each other and removing the Berry curvature from the bands \cite{Rechtsman:13b, Delplace:11}.  A summary of the differences between honeycomb and brick-wall lattice Haldane models can be found in Table \eqref{diff}.

\section{Synthetic brick-wall lattice}
\subsection{From Laboratory to Target Hamiltonian}
\begin{figure}
\includegraphics[draft=false,width=\columnwidth]{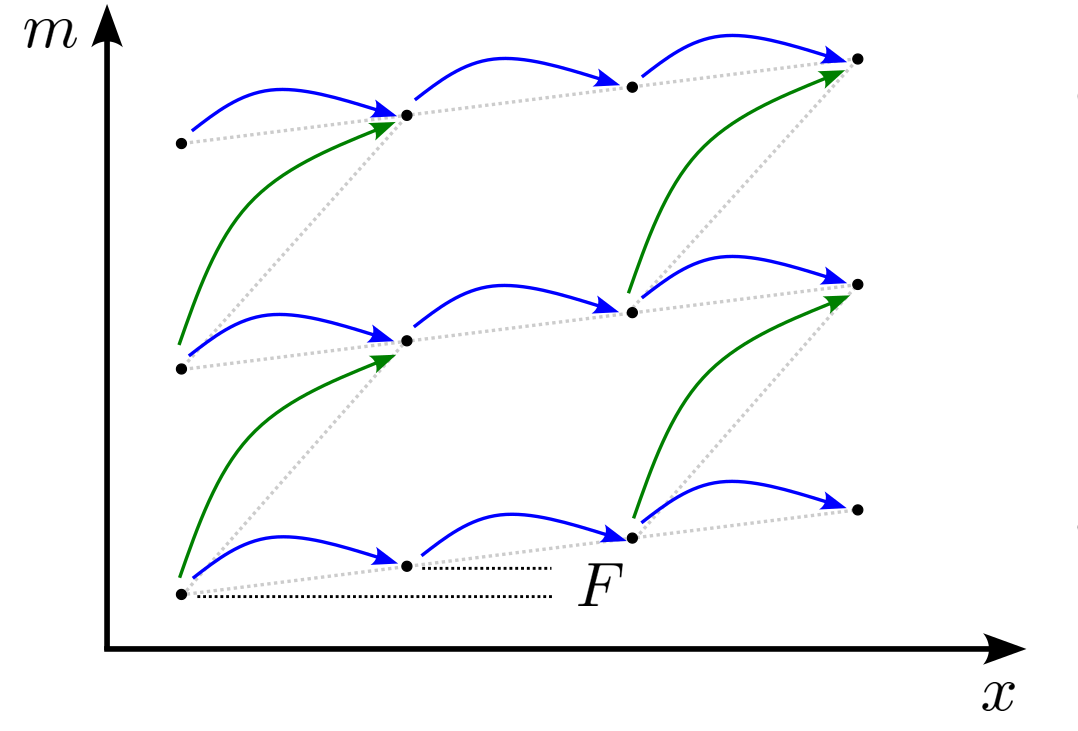}
\caption{Brick-wall lattice structure due to Raman-assisted tunnelling with linear tilt $F$ Arrows indicate positive phase gain}
\label{laser}
\end{figure}

Our proposal has two coupled dimensions, one of which is real and the other is a synthetic dimension constructed from the internal states of the atoms, collectively known as the hyperfine manifold. In the real dimension, which we take to be orientated along the $x-$axis, there is an optical lattice within which resides a species of atom. The optical lattice is constructed using counter-propagating laser beams and we assume strong confinement along other axes, thus limiting real-space dynamics to along the direction of beam propagation. A linear potential with amplitude $F>0$ is applied to the lattice which induces a tilt in energy along $x-$ axis. 
\begin{figure}
\includegraphics[draft=false,width=\columnwidth]{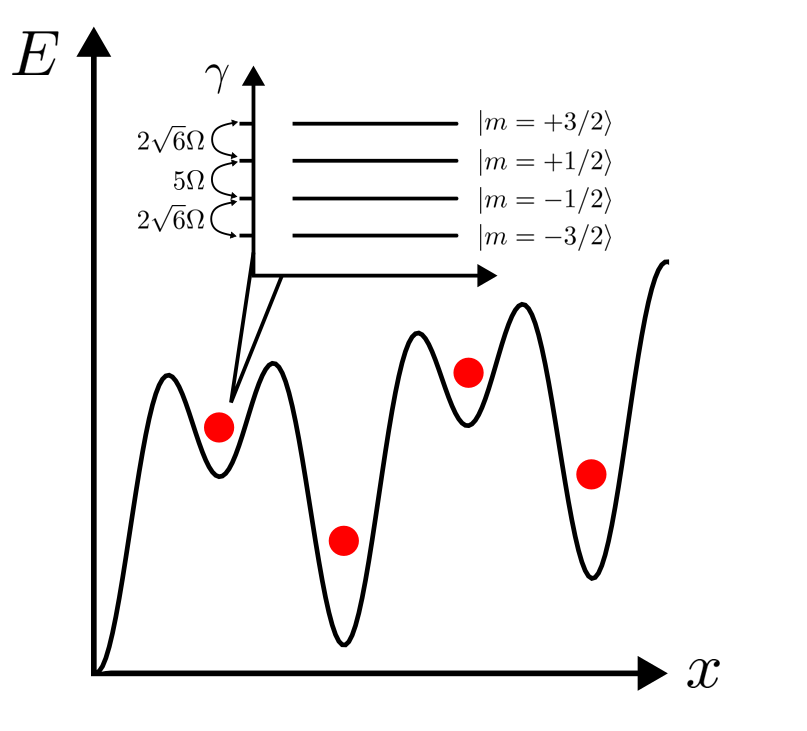}
\caption{Schematic diagram of the system. Red dots represent atoms and the internal states (inset plot) used as an example are the four lowest spin projection hyperfine states from a $F=9/2$ manifold.}
\label{reallattice}
\end{figure}
This creates a laboratory Hamiltonian known as the Wannier-Stark Hamiltonian
\begin{align}\label{WS}
    \hat{H}_{WS} = \frac{\hat{p}^2}{2m}+V\:\text{cos}^2(k_L \hat{x}) +F\hat{x}
\end{align}
where $V$ is the lattice depth and $k_L$ is the wave-vector of the lattice lasers. Now, one must construct the target Haldane Hamiltonian. First, we add to \eqref{WS} a second potential with half the wavelength which makes the lattice bipartite, and therefore the beams form an optical superlattice. Hence, one obtains
\begin{align}\label{WSext}
    \hat{H}_{WS} = \frac{\hat{p}^2}{2m}+V\:\text{cos}^2(k_L \hat{x})+U\:\text{cos}^2(2k_L \hat{x}) +F\hat{x}.
\end{align}\label{secquan}
In second quantised formalism, taking the reasonable approximation of considering only nearest neighbour terms, this becomes
\begin{align}
    \hat{H}_{TB} = &\sum_l (\epsilon_l+aFl)\hat{a}^\dagger_l \hat{a}_l \nonumber \\+ &\sum_l t_1(\hat{a}^\dagger_{l+1} \hat{a}_l+\hat{a}^\dagger_l \hat{a}_{l+1}), 
\end{align}
where 
\begin{align}\label{WSenergy}
    \epsilon_l=&E+V\:\text{cos}^2(k_L a l)+U\:\text{cos}^2(2k_L al)\nonumber\\
    =&E+V+\frac{1}{2}\left[1+(-1)^l\right]U
\end{align}
that is to say the optical lattice inherently introduces an energy difference of $U$ between adjacent sites. We have used, in reaching the final line of Equation \eqref{WSenergy} , $k_L=\frac{\pi}{a}$ and $$E=-\frac{\hbar^2}{2m}\int^\infty_{-\infty} \text{d}x\;\psi^\dagger_l(x)\partial^2_x\psi_l(x),$$ where $\psi_l(x)$ is the Wannier function on site $l$. This is a function constructed from the Bloch functions and the resulting basis is of states exponentially localised to the physical sites \cite{Suszalski:16}
\begin{align}
    \psi_l(x)=\int_{-\pi/2}^{\pi/2}\text{d}k\; e^{-ikla}\phi_k(x)
\end{align}
where $\phi_k(x)$ is the Bloch function with quasimomentum $k$.
The linear potential prevents tunnelling of atoms between neighbouring sites. Additional control is required in the hopping rates to realise the Haldane model together with next-nearest neighbour tunnellings. Nearest neighbour hopping is reintroduced by the use of Raman-assisted tunnelling. The change in energy caused by such a process can be made equal to the energy difference between sites and thus particle hopping is induced. At the same time, a complex phase can be imbued on the atom by having the beam incident on the optical lattice at an angle, which imparts a Raman recoil momentum \cite{Celi:14}. The internal states of the atom can be coupled together, also by the use of Raman beams or by the use of a radio-frequency magnetic field \cite{Stuhl:15}. Nearest neighbour and next-nearest neighbour hoppings in real-space have been experimentally achieved \cite{Miyake:13, Jotzu:14} and for internal states their implementation has been outlined in \cite{Boada:12}. Changing both real and synthetic sites can also be done simultaneously \cite{Suszalski:16} and as will be subsequently shown, due to the `geometry' of couplings that will be used, this will be a requirement. As these amplitudes need not be complex due to translational invariance, one can set $q_x = 0$, and hence $t_{j\rightarrow j+1} = t_1$.
In the tight-binding model described thus far, the atoms reside in the lowest rung of what is known as the Wannier-Stark ladder. The tunnelling rate induced by the Raman-assisted tunneling for nearest neighbours is given by \cite{Miyake:13, Miyake:Thesis, Suszalski:16}
\begin{align}\label{next}
t_{j\rightarrow j+1}=\Omega e^{-ijq_x} \int\text{d}x\; W_0(x)W_1(x) e^{-iq_xx},
\end{align}
where $\Omega$ is the effective Rabi frequency, $j$ is a physical site index and $W_i(x)$ is the real-valued Wannier-Stark function at site $i$. Hence, the complex phase imprinted is given in a site dependent way, $q_x$ is the momentum transfer along the $x-$direction due to the laser and so the phase is $e^{-ijq_x}$.
This allows for the engineering of a synthetic magnetic flux through a plaquette which these complex hoppings bound. Using Raman-assisted tunneling to couple adjacent sites in the optical lattice and coupling adjacent sites in the optical lattice whilst also moving up in internal state, which we will refer to as a diagonal process, one can realise a brick-wall lattice structure. There is one extra step however; one must remove every other diagonal hopping element. This is done by adjusting tunnelling phases such that the two Raman processes add destructively on every other site. This only occurs if the processes have the same amplitudes, see Figure \eqref{laser} for the resulting semi-synthetic lattice where blue arrows indicate tunnelling purely in real space, \textit{i.e.} hopping along the optical lattice, and green arrows represent the diagonal process, absent on every other site. This scheme is due to Ref. \cite{Suszalski:16}.

We go one step further and propose that by starting from this brick-wall structure with all hopping terms phaseless and adding complex next-nearest neighbour terms, one can achieve the Haldane model. To do this we include next-nearest neighbour terms in the second quantised formalism. The phase gain of this hopping is obtained with the directionality included, as arises naturally with the use of Raman-assisted tunnelling
\begin{align}\label{pred}
    \hat{H}_{TB} = &\sum_l (\epsilon_l+aFl)\hat{a}^\dagger_l \hat{a}_l \nonumber \\+ &\sum_{\langle l,m\rangle} t_1(\hat{a}^\dagger_m \hat{a}_l+\hat{a}^\dagger_l \hat{a}_{l+1})\nonumber \\+&\sum_{\langle\langle l,m\rangle\rangle} t_2(e^{i\phi}\hat{a}^\dagger_m \hat{a}_l+e^{-i\phi}\hat{a}^\dagger_l \hat{a}_m),
\end{align}
where $t_1$ and $t_2$ are the nearest neighbour and next-nearest neighbour hoppings rates, respectively. These will directly correspond to the tunnelling coefficients in the target Hamiltonian, Equation \eqref{Haldane}, as will the phase $\phi$. Identifying $M = U/2$ and separating the site operators into those for $A$ and $B$ sites, to allow for spinorial notation, one can then write
\begin{align}\label{TBHaldane}
    \hat{H} = &\sum_n M\left[\hat{\psi}^\dagger_{n,A}\hat{\psi}_{n,A}-\hat{\psi}^\dagger_{n,B}\hat{\psi}_{n,B} \right] \nonumber \\
    + &\sum_{\langle n,m\rangle} t_1 \left[\hat{\psi}^\dagger_{n,B}\hat{\psi}_{m,A}+\hat{\psi}^\dagger_{n,A}\hat{\psi}_{m,B}\right] \nonumber \\
    +&\sum_{\langle\langle l,m\rangle\rangle} t_2 \left[e^{i\phi}\hat{\psi}^\dagger_{n,A}\hat{\psi}_{m,A}+e^{-i\phi}\hat{\psi}^\dagger_{m,A}\hat{\psi}_{n,A}\right] \nonumber \\
    +&\sum_{\langle\langle l,m\rangle\rangle} t_2 \left[e^{i\phi}\hat{\psi}^\dagger_{n,B}\hat{\psi}_{m,B}+e^{-i\phi}\hat{\psi}^\dagger_{m,B}\hat{\psi}_{n,B}\right].
\end{align}

\begin{figure}[!h]
\includegraphics[draft=false,width=\columnwidth]{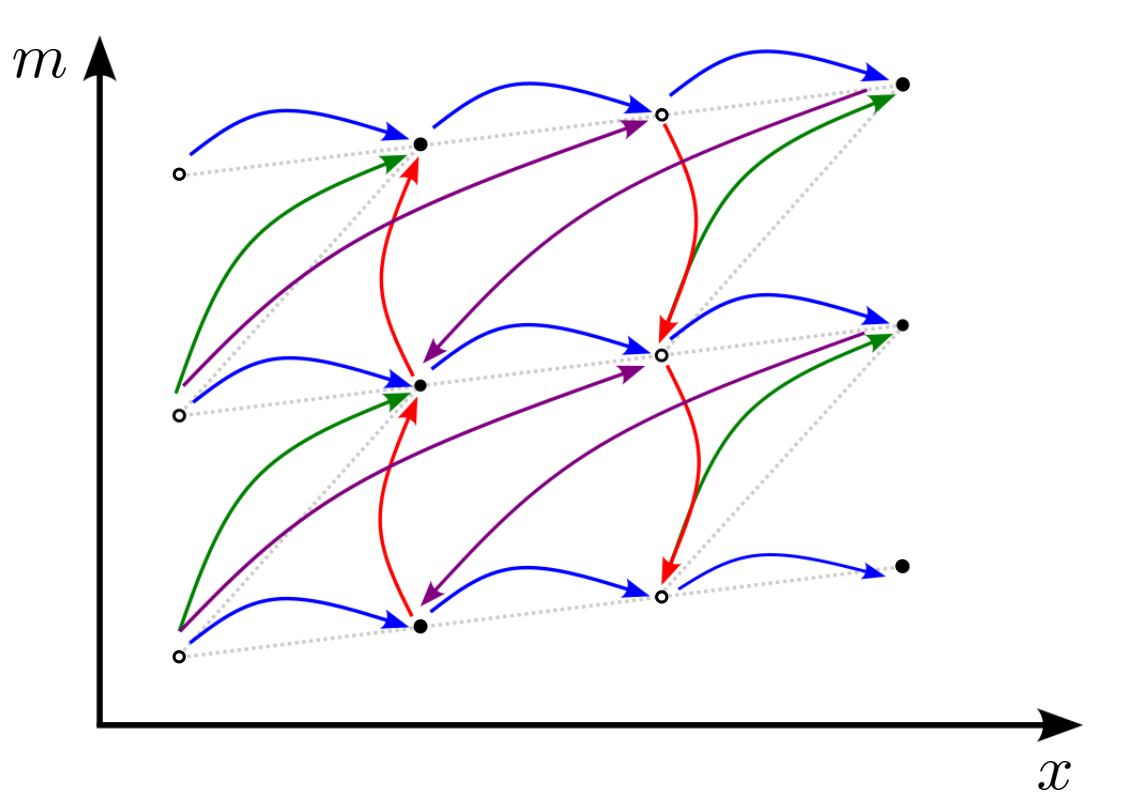}
\caption{Brick-wall lattice structure, conventions as in Figure \eqref{laser}, now with next-nearest neighbour tunnelling. White (black) dots represent $A$($B$) sites, made distinct by an energy offset}
\label{lasernnn}
\end{figure}

The term $aFl$ in Equation \eqref{pred} is experienced by all sites and is not included in the derivation from here on, as it is a general energy shift only.
Fourier transforming Equation \eqref{TBHaldane} changes it into the form of the Bloch Hamiltonian of the Haldane Model in Equation \eqref{Haldane}.
As the synthetic dimension brick-wall lattice constructed is skewed, the next-nearest neighbour hoppings come in two forms. The first type, is in fact, a simple hopping in adjacent internal states with no change in the physical site, and the second is a more involved next-nearest neighbour tunnelling in real space accompanied by a change in internal state, which we will refer to as a `long-diagonal' process, see Figure \eqref{lasernnn}, where purple arrows represent long-diagonal process and red arrows represent a straightforward internal state change. As with the other hopping processes, these may be achieved by Raman-assisted tunnelling, by matching the Raman-frequencies to the energy difference between the desired states. The double hopping in real space, without an internal state change, is therefore achieved by tuning the effective Raman frequency to twice the frequency of fundamental resonance \cite{Miyake:13}. 
Tunnelling rates of at least 1 Hz are required for experimental implementations of models involving next-nearest neighbour atoms \cite{Miyake:13} and this is achievable thanks to Raman-assisted tunneling as the bare tunnelling may be much smaller than this rate. Further to this, an advantage of using Raman-assisted tunnelling is that nearest neighbour and next-nearest neighbour tunnelling rates can be controlled independently. The first experimental realisation of the Haldane model achieved next-nearest tunnelling rates of 5 and 18 Hz. This was done by way of an elliptically modulated honeycomb-like lattice, giving rise to an effective Hamiltonian with complex next-nearest neighbour couplings which the $^\text{40}\text{K}$ atoms experience (\cite{Jotzu:14}; see in particular, Methods and Supplementary Materials).

\subsection{Energy Hierarchy}
Other restrictions on energies do exist, creating a hierarchy of energy scales for the proposal and these are similar to those for the Harper-Hofstadter model \cite{Miyake:Thesis}. For example, the tilt of the potential must be greater than the bare tunnelling coefficient $J$ to inhibit this natural tunnelling, as was discussed at the start of this section. The tilt however must be less than the band gap $\omega$ of the Wannier-Stark model, as groundstates can then occupy excited states of adjacent sites hence, $J < F < \omega$. To ensure the bands of the energy spectrum do not overlap, the ratio of the next-nearest neighbour tunnelling coefficient to that of the nearest neighbour should fulfil $|t_2|<|t_1|/3$.
\newline

As in the Haldane model, the lattice must be bipartite and this is done by the introduction of an additional pair of counter-propagating laser beams with double the wavelength of the original lattice beams, and so increasing the on-site energy of every other site. The optical lattice is now an optical superlattice and the sites can be divided into two types $A,B$, which breaks inversion symmetry, creating a sublattice with an energy offset of $M \neq 0$. This in turn gives two different tunnelling coefficients due to energy differences between sites, depending on the direction, of $\Delta_\pm = F \pm M$, where $M$ is the contribution from the additional laser beams and will be the same $M$ as discussed in previous section. Thus, the hierarchy is modified to $J < \Delta_-\leq \Delta_+ < \omega$. If one wishes to see a topological structure then the energy offset is required to be such that $M < 2 \sqrt{3} t_2 \;\text{sin}(\phi)$. This is for the four next-nearest neighbour case and the argument as to the origin of this inequality is the same as discussed in Section II for the location of the phase transitions.
\newline

Now that the lattice is bipartite, an additional pair of lasers is required, where the Raman frequencies of the two pairs of lasers for real-space hopping must be equal to $\Delta_\pm$. Next-nearest neighbour hoppings, $A-A$ and $B-B$, require a Raman frequency of $2aF$ and so a long diagonal process requires $2aF+\gamma_{m,m+1}$, where $\gamma_{m,m+1}$ is the Zeemann level energy splitting from state $m$ to $m+1$ as there may be an inhomogeneity in these energies. Hence, a change in purely the synthetic dimension will require a Raman frequency of $\gamma_{m,m+1}$. These final two processes need to imprint a complex phase on the wavefunction of the atom. It should be noted that purely real-space complex next-nearest neighbour couplings can be achieved by an additional pair of laser beams, but as discussed in Section II, these are unnecessary to see non-trivial topological responses.

\subsection{Next-nearest neighbour tunnelling phases}
\subsubsection{Synthetic dimension transitions}
The phase which should be imprinted for the two different next-nearest neighbour transitions needs to be carefully chosen. For the purely synthetic transition a phase of $\phi$ should be used. This can be tuned using the momentum transfer, denoted $q_s$, of the Raman beam which itself can be varied by changing the angle of incidence used. 
The tunneling rate for such a transition is given by \cite{Miyake:Thesis}
\begin{align}\label{synth}
t_j=\Omega e^{-ijq_s} \int \text{d}x\; |W_0(x)|^2 e^{-iq_sx} 
\end{align}
This leads to a phase gain of $e^{-ijq_s}$, where $j$ denotes the optical lattice site that the atom left. When an atom makes the following transitions 
$$(j,m)\rightarrow(j,m+1)\rightarrow(j+1,m+1)\rightarrow(j+1,m)$$ 
there is an overall phase gain given by $-\phi j - (-\phi)(j+1) = \phi$, or equivalently there is a synthetic magnetic flux of $ \phi$ through every plaquette. This is not too dissimilar from the two-dimensional real lattice with an effective magnetic field realised in Ref. \cite{Aidelsburger:11}, demonstrating that high effective fields strengths can be attained. See also the work of \cite{Aidelsburger:11, Aidelsburger:13, Atala:14} technical details.

\subsubsection{Long diagonal transitions}
For the other next-nearest neighbour hopping, a long-diagonal process of two real-space hoppings and an internal level change, the required phase is $\phi/3$. When an atom makes the following transitions $$(j,m)\rightarrow(j+2,m+1)\rightarrow(j+3,m+1)\rightarrow(j+1,m)$$ 
there is an overall phase gain given by $-\frac{1}{3}\phi j - (-\frac{1}{3}\phi)(j+3) = \phi$. Again, here for long-diagonal processes, there is an equivalent magnetic lux $\phi$ piercing the plaquette, which in this instance is rhombus shaped.

The tunneling rate for such a transition is given by \cite{Miyake:Thesis}
\begin{align}\label{long}
t_{j\rightarrow j+2}=\Omega e^{-ijq_x} \int \text{d}x\;W_0(x)W_2(x) e^{-iq_xx}.
\end{align}

The number of lasers may be reduced if the required frequency, polarisation and intensity of any two coincide. As discussed for the triangular lattice in Ref. \cite{Suszalski:16}, three tunnellings may be created from three lasers, where a Raman process usually involves two laser beams.

To increase the synthetic dimension length, more states from the hyperfine manifold must be coupled together. The spacing in energy may not be the same between any two adjacent sites, hence the Rabi frequencies need to be adjusted according to Clebsch-Gordan coefficients. The energy spacings are given by $\Omega g_{F,m}$ with $$g_{F,m} = \sqrt{F(F+1)-m(m+1)},$$ where $F$ denotes the quantum angular momentum state number and $m$ denotes the projection of the angular momentum. For alkali metals it has been noted that these modifications are not negligible (for example; see Supplementary Materials \cite{Celi:14}). This means that for every `ribbon' of synthetic dimension one wishes to include in the lattice, extra lasers coupling the extra internal states may be required. 

One would imagine that a minimum of three internal $m$ states are required; one `largest' internal state hosting an edge state with one direction of propagation, a `smallest' internal state hosting an edge state with a current flow opposite to that previously mentioned, and sandwiched between them, an internal state which has minimal particle transport. This was the case for the quantum Hall effect synthetic dimension experiments \cite{Mancini:15, Stuhl:15}. Finite size effects in synthetic dimensions are non-trivial and have been considered previously for the Harper-Hofstadter model, where it has been shown that a topological signature may still be extracted \cite{Genkina:19}.

\section{Edge States}
A good starting point for considering edge states in our proposal is the experimental realizations of the quantum Hall effect in synthetic dimensions. Using a three-leg lattice in a quantum Hall regime \cite{Mancini:15, Stuhl:15}, time-of-flight methods were used to obtain the lattice momentum distribution $n(k)$ by a series of spin-selective images, tracking the expansion of the atomic cloud.
Next, the lattice momentum imbalance, a quantification of the asymmetry in the momentum distribution for a leg $m$, is given by 
$$\mathcal{I}_m = \int \text{d}k_x \;[n_m(k_x)-n_m(-k_x)],$$
and for the top and bottom legs of the lattice (for example, $m = \pm1$ in Ref. \cite{Stuhl:15}), this value was found to be equal and opposite, while for the central leg ($m=0$, \cite{Stuhl:15}) it was found to be close to zero. By using quench dynamic experiments, the classic cyclotron motion can be observed, which gives a skipping-effect along the boundary, as expected. In this proposal the edge states are located on an edge which corresponds to the zig-zag edge. To realise an armchair edge, one would need to exchange the synthetic and real dimensions, which would most likely require highly-involved, spin-dependent and site-dependent tunnelling. Edge states on armchair lattices, however, are only thought to exist in the presence of anisotropy \cite{Kohmoto:10}.

\section{Interactions}
Considered separately, the topics of interactions in synthetic dimensions and in topological systems are fascinating problems. In synthetic dimensional systems, interactions which take place between atoms, where the internal state of the atom is the extra dimension, may give rise to non-local effects since the atoms can reside in the same physical site but separated in synthetic dimension.

For example, consider all the atoms on some same site in Figure \eqref{lasernnn} and notice they are `linked' to one another by the red arrows, which represent the purely synthetic dimensional hopping as discussed in Section III C. 1. . Considering what `arrows' these would be in the brick-wall lattice with the aid of Figure \eqref{lattices} (b), 
one realises that they are \textit{next-}nearest neighbour hoppings and so interactions would take place in the brick-wall lattice \textit{only} along some axis at an angle $\frac{\pi}{4}$ with respect to the lattice vectors. Thus interactions in the synthetic dimensions would simulate highly anistropic interactions in  the original model. Coherent collisions in spin-1 Bose-Einstein condensate has been studied to the extent that it has even been experimentally exploited to determine a spin-dependent scattering length \cite{Chang:05}. In this instance, the internal dynamics are decoupled from the spatial macroscopic wavefunction as the spin healing length is larger than the size of the condensate. Ideally, spin-changing collisions would be suppressed or population dynamics would average nicely to allow for the extraction of topological signatures. 

The Chern number classifies the topology of a gapped band structure for a non-interacting system and so two bands with the same Chern number may be deformed into each other without closing the gap and changing any properties. First we will discuss the spinless case. Weak interactions in the Haldane model have little effect on the system \cite{Hasan:10}. However for repulsive nearest-neighbour interactions of the same order of magnitude as nearest neighbour hopping, the topological phases are destroyed and a charge density wave insulating state appears \cite{Alba:16}. Expanding on this work it has been determined that adding on-site disorder interactions and varying the strengths of the two different types of interactions can give rise to different types of insulating states depending on the relative strengths of the interactions \cite{Yi:21}.

Interactions between electrons are seemingly a prerequisite for the appearance of the fractional quantum Hall effect \cite{Feldman:21}. In a similar vein, interactions have been considered as a route to fractionalisation in ultra-cold atom systems too, by way of charge pumping \cite{Zeng:15}. The infinite range interaction found along a synthetic dimension is a large obstacle which needs to be circumvented to realise a fractional Chern insulator in such a system. These problematic interactions may be made short ranged by way of Trotterization of the system \cite{Barbiero:20} and therefore provide a basis for fractional excitations in synthetic dimensions. Taking a different route, by coupling the top- and bottom-most legs of the synthetic ladder and reinterpreting the system as a cyclinder, it has been shown that fractional quantum Hall states in the form of density waves can arise \cite{Barbarino:15}.

An \textit{attractive} Hubbard interaction, which is again infinite ranged along the synthetic dimension, has been shown to induce Cooper pairing between the counterpropagating chiral edge states in quantum Hall ribbons \cite{Stuhl:15, Mancini:15}, leading to a one-dimensional topological superfluid exhibiting Majorana physics\cite{Yan:15}. Furthermore, exotic states of matter such as the supersolid are found in synthetic models with interactions. For example, a bipartite lattice inhabited by internally-coupled atoms and with next-nearest neighbour hopping, at low tunneling rates there exists; a charge density wave at exactly half-filling, a supersolid on one sublattice below this filling fraction, and phase separation above this filling fraction. At high next-nearest neighbour tunneling rates, there exists a phase which is dominated by superfluid order \cite{Bilitewski:16}. 

In summary, the inclusion of interactions affects the order of the quantum fluid and will therefore have an effect on the model the quantum fluid is simulating.

\section{Conclusion}
We have outlined a proposal for a semi-synthetic dimensional realisation of the Haldane model. Our scheme exploits the fact that the brick-wall lattice is topologically equivalent to the honeycomb lattice. By first realising this in a synthetic fashion by use of an optical lattice and coupled internal states of an atom, one may go on to realise the Haldane model. This is done by the introduction of complex next-nearest neighbour hopping terms via Raman-assisted tunneling and by making the lattice bipartite. The flux through a plaquette of the lattice can be changed by varying the angle of incidence of the Raman beams. One can additionally change the energy offset between adjacent sites of the optical superlattice, giving experimental access to the full range of parameters required to explore the phase diagram of the model. Our proposal could be a new platform for the investigation of topologically non-trivial materials. For example, one could extend the model into the $x-y$ plane where each row of sites displays this model. Coupling this system to a copy of itself would emulate a bi-layer Haldane model which is an interesting problem to consider \cite{Bhattacharjee:21}.

When the energy spectrum exhibits no band gap, for example at $\phi = \frac{\pi}{2}$ and $M = 2\sqrt{3}t_2$, the system is referred to as a topological semi-metal, which may have a non-trivial winding number if the Fermi energy is such that the lower band is filled \cite{Goldman:13a}. Our proposal may allow for a realisation of such states of matter, however, the origin of this phase is subtle, as it is due to the laser recoil momentum which is not a feature of the original Haldane model. The analogous situation with regards to synthetic dimensions is an open question.

Whilst writing this manuscript the authors were made aware of a similar proposal utilising Haldane-like phases of a semi-synthetic dimensional system for the investigation of the generalised Wilson-Dirac model. The Hamiltonians of both systems are very similar in mathematical structure but are distinct due to the fact that the complex-tunneling term is nearest neighbour hopping as opposed to next-nearest neighbour. In the article, Raman-assisted tunneling and the coupling of internal states are also considered in great depth and we encourage readers to compare and contrast our proposals \cite{Kuno:18}.

\section*{ACKNOWLEDGEMENTS}
The authors acknowledge helpful discussions with Stewart Lang. J.P. acknowledges financial support from EPSRC DTP Scholarship and G. V.-R. acknowledges financial support from EPSRC CM-CDT Grant No. EP/L015110/1.

%\bibliography{Bibliography.bib}

\end{document}